\documentclass[9pt,twocolumn,twoside]{pnas-new}
\templatetype{pnasresearcharticle} % Choose template 
\title{Confidence Collapse in a Multi-Household, Self-Reflexive DSGE Model}

\newcommand*{\sic}{{{C_t}}}

\newcommand*{\esp}{{\mathbb{E}}}
\newcommand*{\prof}{{\mathbb{P}}}

\author[a,b,c]{Federico Guglielmo Morelli}
\author[b,c,d,1]{Michael Benzaquen} 
\author[a]{Marco Tarzia}
\author[d,c,e]{Jean-Philippe Bouchaud}

\affil[a]{LPTMC, UMR CNRS 7600, Sorbonne Universit\'e, 75252 Paris Cedex 05, France}
\affil[b]{Ladhyx, UMR CNRS 7646 \&  Department of Economics, Ecole Polytechnique, 91128 Palaiseau Cedex, France}
\affil[c]{Chair of Econophysics and Complex Systems, Ecole polytechnique, 91128 Palaiseau Cedex, France}
\affil[d]{Capital Fund Management, 23-25, Rue de l'Universit\'e 75007 Paris, France}
\affil[e]{Acad\'emie des Sciences, Quai de Conti, 75006 Paris, France}

\leadauthor{F. G. Morelli} 

\correspondingauthor{\textsuperscript{1}To whom correspondence should be addressed. E-mail: michael.benzaquen@polytechnique.edu}

\keywords{
\textbf{Significance statement:} Despite their inability to cope with the Global Financial Crisis and various subsequent calls to ``Rebuild Macroeconomics'', Dynamics Stochastic General equilibrium (DSGE) models are still at the forefront of monetary policy around the world. Like many standard economic models, DSGE models rely on the figment of representative agents, abolishing the possibility of genuine collective effects (such as crises like the 2008 GFC) induced by heterogeneities and interactions. By allowing feedback of past aggregate consumption on the sentiment of individual households, we pave the way for a class of more realistic DSGE models that allow for large output swings induced by relatively minor variations in economic conditions and amplified by interactions. Several important conceptual messages follow from our work, for example the {\it de facto} impossibility to price extreme risks and the importance of narratives, which may be an efficient depression-prevention policy tool whenever confidence collapse is looming.
} 

\begin{abstract}
We investigate a multi-household DSGE model in which past aggregate consumption impacts the confidence, and therefore consumption propensity, of individual households. We find that such a minimal setup is extremely rich, and leads to a variety of realistic output dynamics: high output with no crises; high output with increased volatility and deep, short lived recessions; alternation of high and low output states where relatively mild drop in economic conditions can lead to a temporary confidence collapse and steep decline in economic activity. The crisis probability depends exponentially on the parameters of the model, which means that markets cannot efficiently price the associated risk premium. We conclude by stressing that within our framework, {\it narratives} become an important monetary policy tool, that can help steering the economy back on track.   
\end{abstract}

\dates{This manuscript was compiled on \today}
\doi{}

\begin{document}

\maketitle
\thispagestyle{firststyle}
\ifthenelse{\boolean{shortarticle}}{\ifthenelse{\boolean{singlecolumn}}{\abscontentformatted}{\abscontent}}{}
\section{Introduction} 

In spite of their poor performance during the Global Financial Crisis (GFC), Dynamics Stochastic General Equilibrium (DSGE) models still constitute the workhorse of monetary policy models around the world (see e.g. \cite{gali2015monetary} for an insightful introduction and references). Many ingredients that were missing in previous versions of the model (such as the absence of a financial sector) have been added in the recent years, in an attempt to assuage some of the scathing criticisms that were uttered post GFC (see for example \cite{buiter2009unfortunate,trichet2010reflections,blanchard2014danger,stiglitz2018modern} and \cite{christiano2018dsge, blanchard2018future, reis2018something} for rebuttals). 
However, the whole DSGE framework seems to be -- partly for technical reasons -- wedded to the Representative Agent (or Firm) paradigm, and to a (log-)linear approximation scheme that describes small perturbations away from a fundamentally stable stationary state.\footnote{Quoting O. Blanchard in \cite{blanchard2014danger}: {\it We in the field did think of the economy as roughly linear, constantly subject to different shocks, constantly fluctuating, but naturally returning to equilibrium over time. [...]. The problem is that we came to believe that this was indeed the way the world worked.}} In other words, crises are difficult to accommodate within the scope of DSGE, a situation that led to J.C. Trichet's infamous complaint: {\it Models failed to predict the crisis and seemed incapable of explaining what was happening [...], in the face of the crisis we felt abandoned by conventional tools} \cite{trichet2010reflections}.  

Agent Based Models (ABMs) provide a promising alternative framework to think about macroeconomic phenomena \cite{gatti2008emergent,fagiolo2012macroeconomic,gualdi2015tipping,hommes2018computational,dosi2019more}. In particular, ABMs easily allow for heterogeneities and interactions. These may generate non-linear effects and unstable self-reflexive loops that are most likely at the heart of the 2008 crisis, while being absent from benchmark DSGE models where only large technology shocks can lead to large output swings. Unfortunately, ABMs are still in their infancy and struggle to gain traction in academic and institutional quarters (with some major exceptions, such as the BoE \cite{haldane2018interdisciplinary} or the OECD). In order to bridge the gap between DSGE and ABMs and allow interesting non-linear phenomena, such as trust collapse, to occur within DSGE, we replace the representative household by a collection of homogeneous but interacting households. Interaction here is meant to describe the feedback of past aggregate consumption on the {\it sentiment} (or confidence) of individual households -- i.e. their future consumption propensity. Low past aggregate consumption begets low future individual consumption. This opens the possibility that a relatively mild drop in economic conditions leads to a confidence collapse and a steep decline in economic activity.

\begin{figure*}[t!]
\centering
\includegraphics[width=\textwidth]{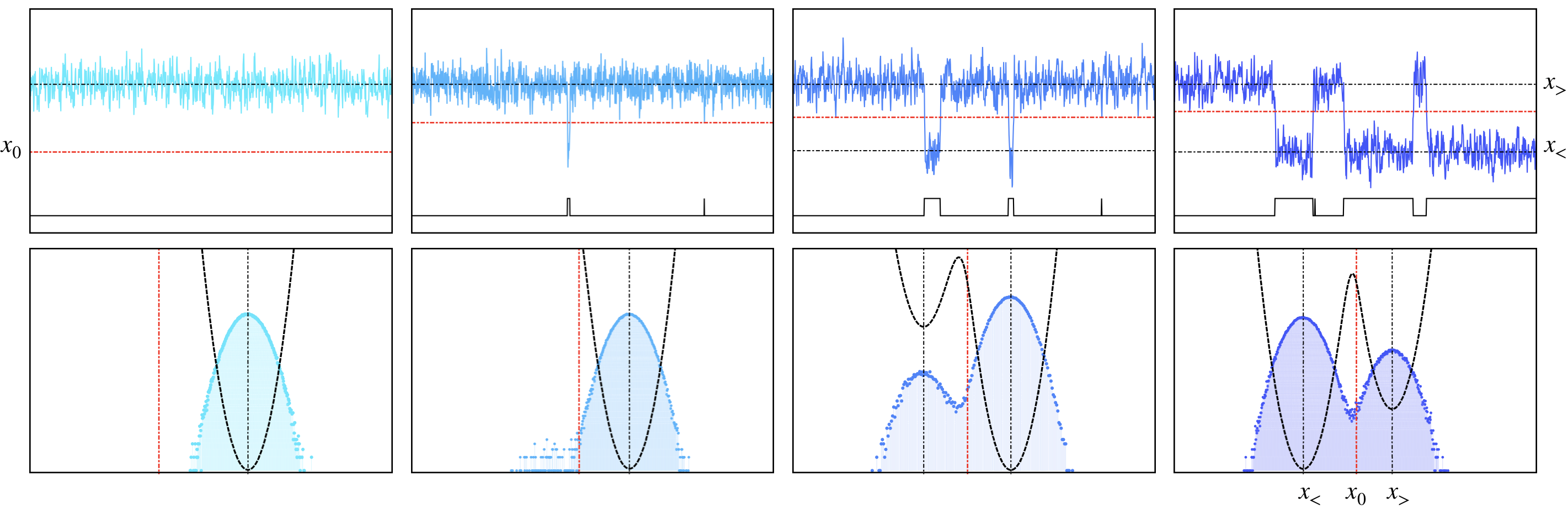}
\vspace{-0.5cm}
 \caption{Numerical simulation of the model for increasing values of the confidence threshold $c_0$ and for fixed values of $\theta= 5 $, $\sigma= 0.6$ and $\eta=0.5$. Top graphs: temporal trajectories of the log output $x_t := \log c_t$ with a horizontal dot-dashed red line located at $x_0$ and dashed black lines at $ x_{>,<}$; Bottom graphs: (log-)probability distribution $p(x)$ of the output, with the corresponding positions of $x_0$ and $x_{>,<}$. From left to right: $e^{x_0}=0.1$ ($A$ phase, no crises, Gaussian distribution of output); $e^{x_0}=0.55$ ($B^+$ phase, short crises, increased volatility and skewed distribution of output); $e^{x_0}=0.75$ ($C$ phase, long recessions, bi-modal distribution with most weight on $x_>$); $e^{x_0}=1.05$ ($C$ phase, long recessions, bi-modal distribution with most weight on $\log x_<$). Dashed lines: effective potential $2V(x)/\sigma^2$, defined in Sec.~\ref{transition_rates}.
 }
 \label{fig:dynamics}
\end{figure*}

We establish the ``phase diagram'' of our extended model and identify regions where crises can occur. When the feedback effect is weak enough, standard DSGE results are recovered. As the strength of feedback increases, the economy can undergo rare, short-lived crises, where output and consumption plummet but quickly recover. For even larger feedback, mild technology shocks can induce transitions between a high output state and a low output state in which the economy can linger for a long time. In such conditions, output volatility can be much larger than the total factor productivity. As portrayed by B. Bernanke \cite{bernanke1994financial}, this is a ``small shocks, large business cycle'' situation. We show that these endogenous crises exist even when the amplitude of technology shocks is vanishingly small. But in this limit the probability for such crises is {\it exponentially small} and hence, we argue, unknowable and un-hedgeable. Our model thus provides an interesting example of unknown-knowns, where the crisis is a possible state of the world, but its probability fundamentally un-computable.   

Our work relates to various strands of the literature that emphasises the role of multiple equilibria and self-fulfilling prophecies, in particular the work of Brock \& Durlauf on social interactions \cite{brock2001discrete}. Technically, our modelling strategy is akin to the ``habit formation'' or ``keeping up with the joneses'' (KUJ) literature \cite{campbell1999habit,gali1994keeping,stracca2005keeping}, although our story is more about keeping ``down'' with joneses (KDJ), as we are more concerned by self-reflexive confidence collapse than by consumption sprees (which also exist in our model, but in a regime that does not seem to be empirically relevant). We also specifically study the interplay between the co-existence of two stable equilibria and technology shocks, that can induce sudden transitions between these equilibria (i.e. crises). 

Several other scenarii can lead to the coexistence of static equilibria, corresponding to high/low confidence \cite{anand2013epidemics, da2015sudden, Farmer2019behaviour}, high/low output \cite{gualdi2015tipping, bouchaud2017optimal, carlin2018stagnant}, or high/low inflation expectations \cite{de2011animal, bouchaud2017optimal}, trending/mean-reverting markets \cite{chiarella1992dynamics, lux1999scaling, wyart2007self, majewski2018co}, etc. with possible sudden shifts between the two. Multiple equilibria can be either a result of learning from past events, or from strong interactions between individual agents (direct or mediated by markets) -- for a review, see e.g. \cite{bouchaud2013crises}. Another, distinct line of research explores the consequences of having an indeterminate equilibrium, i.e. a stationary solution around which small fluctuations can develop without being pinned by initial conditions (for a recent review and references, see \cite{farmer2019indeterminacy}). These fluctuations are not related to any real economic driving force, but rather the result of self-fulfilling prophecies. In our present model, fluctuations {\it are} triggered by real technology shocks, but are then amplified by a self-reflexive mechanism. Nothing would prevent, however, the existence of further indeterminacy around different stationary points. 

%The outline of this paper is as follows. In Section~\ref{multi} we present the model and its homogeneous solutions. In Section~\label{animal} we introduce the self-reflexive mechanism and present the Phase Diagram of possible outputs of the model in Section~\ref{phase}. In Section~\ref{transition_rates}, we investigate the dynamics. In Section~\ref{inflation} we discuss some perspectives on inflation and monetary policy and conclude in Section~\ref{ccl}.

\section{A Multi-Household DSGE Model} 
\label{multi}

We assume that each household $i\in [\![1,M]\!]$ is characterised by a utility function $U_i(c_t^i, n_t^i)$ that depends on its (unique good) consumption $c_t^i$ and amount of labour 
$n_t^i$ as: 
\begin{equation}
U_i(c^i_t,n^i_t) = f^i_t \frac{(c_t^i)^{1-\varsigma}}{1-\varsigma} - \gamma_i\frac{(n_t^i)^{1+\phi}}{1+\phi} \ ,
\label{eq:utilityinf}
\end{equation}
where $\gamma_i$ is a factor measuring the disutility of labour, and $\varsigma\in~]0, 1]$ and $\phi > 0$ are two $i$-independent parameters such that the utility function has the correct concavity. Standard choices are $\varsigma=1$ (log-utility of consumption) and $\phi=1$. The quantity $f^i_t$ is a time-dependent factor measuring the confidence of household $i$ at time $t$, and hence their propensity to consume. This ``belief function'' \cite{farmer2019indeterminacy} will be responsible for the possible crises in our model (see below).

Each infinitely lived household maximises its future expected discounted utility with a discount factor $\beta < 1$, subject to the budget constraint:
\begin{equation}
p_t c^i_t + \frac{B^i_t}{1+r_t} \leq w_t n^i_t + B^i_{t-1},
   \label{eq:budget}
\end{equation}
where $p_t$ is the price of the good, $w_t$ the wage (assumed to be identical for all households) and $B_i^t$ the amount of bonds paying $1$ at time $t+1$, purchased at time $t$ at price $(1+r_t)^{-1}$, where $r_t$ is the one-time period interest rate (set by the central bank). The maximisation is achieved using the standard Lagrange multipliers method over the quantities $c^i_t$, $n^i_t$ and $B^i_t$. This gives the household's state equations \eqref{eq:state1}, \eqref{eq:state2} and the Euler equation \eqref{eq:euler}:
\begin{eqnarray}
     \left(c_t^i\right)^{\varsigma} &=& \frac{f^i_t}{\lambda_i p_t} \label{eq:state1} \\
     \left(n_t^{i}\right)^{\phi} &=& u_t p_t \frac{\lambda_i}{\gamma_i} \label{eq:state2} \\
    f_{t}^i \left(c_t^{i}\right)^{-\varsigma} &=& \beta(1+r_t)\esp_t\left[\frac{f_{t+1}^i\left(c_{t+1}^{i}\right)^{- \varsigma}}{1+\pi_{t+1}}\right] \ , \label{eq:euler}
\end{eqnarray}
where $u_t = {w_t}/{p_t}$ is the wage expressed in price units, $\pi_t := {p_t}/p_{t-1}-1$ the inflation rate and $\lambda_i$ the Lagrange multiplier. The total consumption is $\sum_{i=1}^M c_t^i := \sic$ and the total number of work hours is $\sum_{i=1}^M n_t^i := N_t$. 

The unique firm has a technology such that its production $Y_t$ is given by:
\begin{equation}
Y_t = M^\alpha z_t \frac{N_t^{1-\alpha}}{1-\alpha} \label{eq:production},
\end{equation}
where $z_t$ is the total factor productivity and $\alpha$ another parameter, often chosen to be $1/3$. The scaling factor $M^\alpha$ is there to insure a correct limit when $M \to \infty$ that allows market clearing, i.e. total production and total consumption must be both proportional to the number of households $M$. We will write $z_t:= \bar{z}e^{\xi_t}$, where the log-productivity $\xi_t$ is assumed to follow an AR(1) process:
\begin{equation}\label{eq:technology}
    \xi_t = \eta \xi_{t-1} + { \sqrt{1 - \eta^2}}\, \mathcal{N}\left(0, \sigma^2\right),
\end{equation}
where $\eta$ modulates the temporal correlations of the technology shocks, and $\sigma$ the amplitude of these shocks.   

Each time period the firm maximises its profit with the assumption that markets will clear, i.e. that $Y_t \equiv \sic$. Profit is given by $\prof_t:=p_t C_t - w_t N_t$. Maximisation of $\prof_t$ yields $u_t = z_t (M/N_t)^{\alpha}$, i.e., the firm hires labour up to the point where its marginal profit equals the real wage \cite{gali2015monetary}. Now, assuming for simplicity that $f^i_t$ and $\gamma_i$ are 
all equal (homogeneous beliefs and preferences) leads to $c^i_t=c_t=C_t/M$, $n^i_t=n_t=N_t/M$, $\gamma_i=\gamma$ and $f^i_t=f_t$. We can use Eqs. (\ref{eq:state1},\ref{eq:state2}) and Eq. (\ref{eq:production}) with $Y_t=C_t$ to find $c_t, n_t$ and $u_t$ as a function of $f_t$ and $z_t$.  In the following, we choose standard values $\phi=\varsigma=1$, $\alpha=1/3$, yielding\footnote{Other values can of course be considered as well, but do not change the qualitative conclusions of the paper.}
\begin{eqnarray}
    c_t = z_t \left(\frac{9f_t}{4\gamma}\right)^{1/3}. \label{eq:solution}
\end{eqnarray}

\section{Animal Spirits and Self-Reflexivity}
\label{animal}

Now, the main innovation of the present work is to assume that the sentiment of households at time $t$ (which impacts their consumption propensity $f_t$) is a function of the past realised consumption {\it of others}, in an ``animal spirits'' type of way. If household $i$ observes that other households have reduced their consumption in the previous time step, it interprets it as a sign that the economy may be degrading. This increases its precautionary savings and reduces its consumption propensity. Conversely, when other households have increased their consumption, confidence of household $i$ increases, together with its consumption propensity.\footnote{In this work, we assign confidence collapse to households, i.e., to the demand side. However, one could argue that confidence collapse in 2008 initially affected the supply side. One can of course easily generalise the present framework to account for such a ``wait and see'' effect as well. For an other attempts to include behavioural effects in DSGE, see \cite{gabaix2016behavioral}.} A general specification for this animal spirits feedback is 
\begin{eqnarray}
    f^i_t \longrightarrow F\left(\sum_{j=1, j \neq i}^M J_{ij} c^j_{t-1}\right) \label{eq:feedback},
\end{eqnarray}
where $F(.)$ is a monotonic, increasing function and $J_{ij}$ weighs the influence of the past consumption of household $j$ on the confidence level of $i$. In this work, we will consider the case where $J_{ij}=J/M$, i.e. only the 
aggregate consumption matters.\footnote{In the following, we will consider the large $M$ limit such that $J/M \sum_{j=1, j \neq i}^M c^j_{t-1} \to Jc_{t-1}$.  Extensions to non-homogeneous situations will be considered in a future work, but many features reported here will also hold in these cases as well.} This corresponds to a {\it mean field} approximation in statistical physics, see e.g. \cite{kadanoff2009more}. While it neglects local network effects, it captures the gist of the mechanism we want to illustrate and furthermore allows us to keep the household homogeneity assumption (different local neighbourhoods generally lead to different consumption propensities).

\begin{figure}[t!]
\centering
\includegraphics[width = .95\columnwidth]{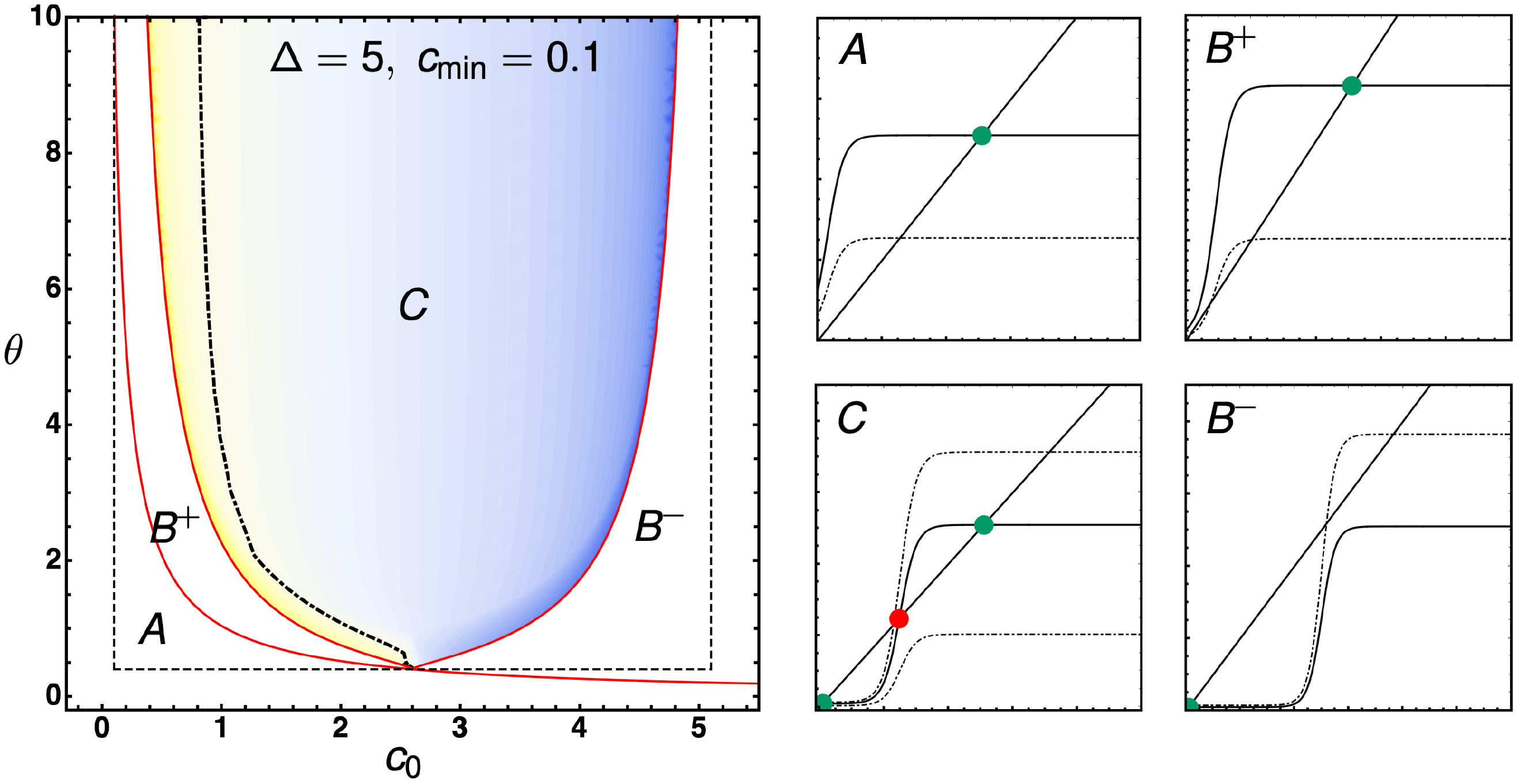}
\vspace{-0.2cm}
\caption{{Left}: Phase diagram of the model, with analytically determined boundaries. Phase $A$: High Output, No Crises; Phase $B^+$: High Output with Short-Lived Recessions; Phase $C$: Long-Lived Booms \& Recessions; Phase $B^-$: Phase $B^-$: Low Output with Short-Lived Spikes. The colour level encodes the distance ratio $(c_>-c^*)/(c^*-c_<)$. This ratio is large in the yellow region, small in the blue region and equal to one along the dot dashed black line. {Right}: Graphical representation of the iteration $c_{t+1}=e^{\xi_t} G(c_t)$ in the different phases. The plain line corresponds to $\xi_t=0$.}
\label{fig:parameter_space}
\end{figure}

Combining \eqref{eq:solution} and \eqref{eq:feedback} yields:
\begin{eqnarray}
    c_t = e^{\xi_t} G(c_{t-1}), \quad \text{with} \quad G(x):=\bar{z} \left(\frac{9F(x)}{4\gamma}\right)^{1/3}. \label{eq:iteration}
\end{eqnarray}
Equation (\ref{eq:iteration}) is a discrete time evolution equation for the consumption level. In order to exhibit how this dynamics can generate excess volatility and endogenous crises, 
we assume that $G(x)$ is a shifted logistic function (but it will be clear that the following results will hold for a much larger family of functions). To wit:
\begin{equation}\label{eq:logistic}
     G(c)= c_{\min} + \frac{c_{\max} - c_{\min}}{1 + e^{2\theta(c_0 - c)}}\, ,
 \end{equation}
 where $c_{\min},c_{\max},c_0$ and $\theta$ are parameters with the following interpretation: 
 \begin{itemize}
 \item $c_{\min} > 0$ is the minimum level of goods that households will ever consume when productivity is normal (i.e. $\xi_t=0$).
 \item $c_{\max} > c_{\min}$ is the maximum level of goods that households will ever consume when productivity is normal (i.e. $\xi_t=0$).
 \item $c_0$ is a ``confidence threshold'', where the concavity of $G(c)$ changes. Intuitively, $c > c_0$ tends to favour a high confidence state and $c < c_0$ a low confidence state.
 \item $\theta > 0$ sets the width over which the transition from low confidence to high confidence takes place:
 %\textcolor{red}{and is proportional to the households' sensitivity to the consumption changes}: 
 in the limit $\theta \to +\infty$, one has $G(c < c_0)=c_{\min}$ and $G(c > c_0)=c_{\max}$.\footnote{In this work we shall fix $\theta$ and vary $c_0$. Note however that fixing $c_0$ and varying the ``temperature" $\theta$ would also be of interest to investigate the effect of population heterogeneity. As mentioned earlier, we leave the study of household heterogeneity for a future work.}
 \end{itemize}
 The standard DSGE model, where the animal spirit feedback is absent, is recovered in the limit $\theta c_0 \to - \infty$, in which case $G(c) \equiv c_{\max}=cst$. 
The dynamics of our extended model falls into four possible phases, that we will call $A$, $B^+$, $C$ and $B^-$ (see Figs. \ref{fig:dynamics}, \ref{fig:parameter_space}), and discuss their properties in turn. In the following, we will use the notation $\Delta:=c_{\max} - c_{\min}$.

This phase corresponds to the DSGE phenomenology, where the only solution of $e^\xi G(c)=c$ is a high consumption solution $c > c_0$ for all values of $\xi$. Even for large negative shocks $\xi < 0$, the economy remains is a confident state. For small noise amplitude $\sigma \ll 1$, the consumption remains around the value $c_>$ solution of $G(c_>)=c_>$, and one can linearise the dynamics around that point: 
\begin{equation}\label{eq:delta}
     \delta_{t+1} \approx G'(c_>) \delta_t + \xi_t, \qquad \delta_t:=\frac{c_t - c_>}{c_>}.
 \end{equation}
This leads to the following expression for the consumption volatility:
\begin{equation}\label{eq:variance}
    \mathbb{V}[\delta] = \frac{\sigma^2}{1 - G_>^{'2}} \frac{1 + \eta G'_>}{1 - \eta G'_>}, \qquad G'_>:=G'(c_>).
 \end{equation}
 In other words, the output volatility is proportional to the amplitude $\sigma$ of the technology shocks -- small shocks lead to small volatility (note that $G'_> < 1$ in the whole $A$ phase). However, the feedback mechanism leads to {\it excess volatility}, since as soon as $G'_> > 0$, one has
 \begin{equation}\label{eq:variance2}
    \frac{\sigma^2}{1 - G_>^{'2}} \frac{1 + \eta G'_>}{1 - \eta G'_>} > \sigma^2.
 \end{equation}

\section{Phase Diagram}
\paragraph{Phase $A$: High Output, No Crises} 
\label{phase}
 
The relative position of the boundary of the $A$ phase depends on whether $\theta \Delta$ is larger or smaller than $2$. In the first case -- corresponding to Fig. \ref{fig:parameter_space}, %(\red{peut-etre representer aussi les autres situation en Fig. 1?})
 the boundary is with the $B^+$ phase (to be described below). In the $(c_0,\theta)$ plane, the $A$ phase is located on the left of the hyperbola defined by
\begin{equation}
    \theta c_0 = 1 + \frac{2 c_{\min}}{\Delta}.
 \end{equation} 
In the case $\theta \Delta < 1$, the boundary is with the $B^-$ phase and the $A$ phase corresponds to $c_0 \leq c_{\min} + \Delta/2$. 

\paragraph{Phase $B^+$: High Output with Short-Lived Recessions} 

%\red{J'ajouterais une figure avec la construction geometrique, meme si on fini pas ne pas la mettre dans le papier a la fin elle sera bien utile pour les presentations.}
 In this phase $B^+$  there is still a unique equilibrium state when productivity is normal, i.e. a unique solution to $G(c_>)=c_>$ with $c_> > c_0$. However, downward fluctuations of productivity can be strong enough to give birth to two more solutions $c_< < c^* < c_0$, one unstable ($c^*$) and one stable ($c_<$). With some exponentially small probability when $\sigma \to 0$ (see \eqref{eq:activated} below), the economy can be driven out of the normal state $c_>$ and crash into a low output state, in which it will remain trapped for a time of the order of $T_\eta := 1/|\ln(\eta)|$, i.e. the auto-correlation time of $\xi_t$. 
 %\textcolor{red}{Are we 100 \% sure of that? I understand this result in the continuum time limit and also intuitively, but I don't know what is the status of Federico's simulations in the discrete setting compared to the continuum tile limit for $\eta$ close to $1$ with the new factor $\sqrt{1 + \eta^2}$.}
 In other words,  sufficiently large fluctuations of output are initially triggered by a relatively mild drop of productivity which is then amplified by the self-referential ``panic'' effect. But since the low output state is only a transient fixed point, the recession is only short lived.

\paragraph{Phase $C$: Long-Lived Booms \& Recessions} 

%\begin{SCfigure*}[\sidecaptionrelwidth][t]
%\centering
%\includegraphics[width=11.4cm,height=11.4cm]{frog}
%\caption{This caption would be placed at the side of the figure, rather than below it.}\label{fig:side}
%\end{SCfigure*}

Phase $C$ is such that equation $G(c)=c$ has two stable solutions $c_<, c_>$ and one unstable solution $c^*$. This phase is delimited, in the $(c_0,\theta)$ plane, by a parabolic boundary (see Fig. \ref{fig:parameter_space}) with $c_0 \to (c_{\min} + c_{\max})/2$ when $\theta \Delta \to 2^+$ and $c_0 \to c_{\min}$ or $c_{\max}$ when $\theta \to \infty$.
%the following asymptotic behaviours (note that there are two $c_0$ solutions for all $\theta>2/\Delta$) \red{\red{**}equations to be checked again\red{**}}:% \red{since we have the exact analytical solution, we can write it and then show the asymptotics in $\theta^{-1}$? Or put in an appendix?}
%\begin{eqnarray}\label{eq:boundary}
%c_0 &=& \frac{c_{\min} + c_{\max}}{2} \pm \frac{\Delta}{12 \sqrt{2}} \left(\theta\Delta - 2\right)^{3/2}, \qquad(\theta \Delta \to 2^+)\nonumber\\
%c_0 &=& c_{\min} + \frac{\log \theta}{2 \theta} + O(\theta^{-1}), \qquad(\theta \to +\infty) \nonumber\\ 
%c_0 &=& c_{\max} - \frac{\log \theta}{2 \theta} + O(\theta^{-1}), \qquad (\theta \to +\infty).
%\end{eqnarray}
The lower boundary $C \to B^+$ corresponds to $c_< \to c^*$ before both disappear, leaving $c_>$ as the only solution, whereas the upper boundary $C \to B^-$ corresponds to $c_> \to c^*$ before both disappear, leaving now $c_<$ as the only solution.

In the absence of fluctuations ($\sigma=0$), the economy in phase $C$ settles either in a low output state or in a high output state. But any, however small, amount of productivity fluctuations is able to induce transitions between these two states. The time needed for such transitions to take place is however {\it exponentially long} when $\sigma \to 0$: 
\begin{eqnarray}\label{eq:activated}
    \log T(c_{>,<} \to c_{<,>}) &=& \frac{W(c_{>,<} \to c_{<,>})}{\sigma^2} + O(\sigma^0);\quad\quad %\nonumber 
 \end{eqnarray}
where $W(c_> \to c_<)$ and $W(c_< \to c_>)$ are computable quantities (see section \ref{transition_rates} below and Fig. \ref{fig:noiseplot}). This is clearly the most interesting regime: the economy can remain for a very long time in a high output state $c_>$, with relatively mild fluctuations (in fact still given by Eq. (\ref{eq:variance})), until a self-fulfilling panic mechanism throws the economy in a crisis state where output is low ($c_<$). This occurs with a Poisson rate $1/T(c_> \to c_<)$. Unless some explicit policy is put in place to restore confidence, the output will linger around $c_<$ for a Poisson time $\sim T(c_< \to c_>)$ which is also very long when $\sigma \to 0$.\footnote{Note in particular that $T(c_< \to c_>)$ is much longer than $T_\eta$: it is no longer the correlation time of productivity fluctuations that sets the duration of recessions (at variance with the $B^+$ scenario).} Note that $T(c_> \to c_<)$ is the average time the system remains around $c_>$ before jumping to $c_<$. The actual time needed to transit is itself short, and the resulting dynamics is made of jumps between plateaus -- see Fig.~\ref{fig:dynamics}. A downward jump therefore looks very much like a ``crisis''. 

As we discuss below, recession durations are much shorter than the time between successive crisis when $c^*-c_< < c_> - c^*$, i.e. when the low output solution is close to the unstable solution, which plays the role of an escape point (see below). As $c_0$ grows larger, one will eventually be in a situation where $c^*-c_< > c_> - c^*$, in which case recession periods are much longer than boom periods, see Fig.~\ref{fig:parameter_space}.    %\red{A figure with the geometric construction would also help illustrate this}
As $\sigma$ grows larger, the output flip-flops between $c_<$ and $c_>$ at an increasingly faster rate, see Fig.~\ref{fig:noiseplot}. While it becomes more and more difficult to distinguish crisis periods from normal periods, the output volatility is dramatically amplified by the confidence feedback loop. 

\paragraph{Phase $B^-$: Low Output with Short-Lived Spikes} 

Phase $B^-$ is the counterpart of phase $B^+$ when $c_0$ is to the right of the phase boundary. In this case, the only solution to $G(c)=c$ is $c_<$: confidence is most of the time low, with occasional output spikes when productivity fluctuates upwards. These output peaks are however short-lived, and again fixed by the correlation time $T_\eta$.\footnote{Note that there is no ``$A^-$'' analogue of the $A$ phase described above -- this is due to the fact that the productivity factor $z_t$ can have unbounded upwards fluctuations but cannot become negative.}  

\paragraph{Phase Diagram: Conclusion}

Although quite parcimonious, our model is rich enough to generate a variety of realistic dynamical behaviour, including short-lived downturns and more prolonged recessions, see Fig. \ref{fig:dynamics}. We tend to believe that the most interesting region of the phase space is in the vicinity of the $B^+$- $C$ boundary, and that the 2008 GFC could correspond to a confidence collapse modelled by a sudden $c_> \to c_<$ transition.\footnote{The role of trust in the unravelling of the 2008 crisis is emphasised in the fascinating book of Adam Tooze, {\it Crashed}, Penguin Books (2018).} The behaviour of the economy in the $B^-$ phase, on the other hand, does not seem to correspond to a realistic situation. One of our most important result is that the crisis probability is exponentially sensitive to the parameters of the model. We dwell further on this feature in the next section.      

\section{A Theory for Transition Rates}
\label{transition_rates}

\paragraph{Discrete Maps} 

Let us now discuss in more detail one of the most important predictions of our model, namely the exponential sensitivity to $\sigma$ of the crisis probability, Eq. (\ref{eq:activated}). Such a result can be obtained by adapting the formalism of \cite{demaeyer2013trace} to the present problem. In terms of $x_t:=\log c_t$, the map (\ref{eq:iteration}) reads:
\begin{equation}\label{eq:gaspard}
    x_t = H(x_{t-1}) + \xi_t,
 \end{equation}
with $H(x):=\log G(e^x)$. In the limit of white noise (i.e. $\eta=0$ in Eq. (\ref{eq:technology})), this is precisely the general problem studied in~\cite{demaeyer2013trace} in the case where $H(x)=x$ has two stable solutions and an unstable one in-between. The authors show that the average time before jumping from one stable solution to another is given, for small $\sigma$, by Eq. (\ref{eq:activated}). They provide an explicit scheme to compute (at least numerically) the quantity $W$, called the {\it activation barrier} in physics and chemistry. The idea is to find the most probable configuration of $\xi_t$'s that allows the system to move from one stable position to another. In a nutshell, this amounts in finding a heteroclinic connection, in an enlarged space, between the starting point and the intermediate, unstable fixed point $x^*=\log c^*$ \cite{gruckenheimer1983nonlinear}.
 
\begin{figure}[b!]
\centering
\includegraphics[width=1\columnwidth]{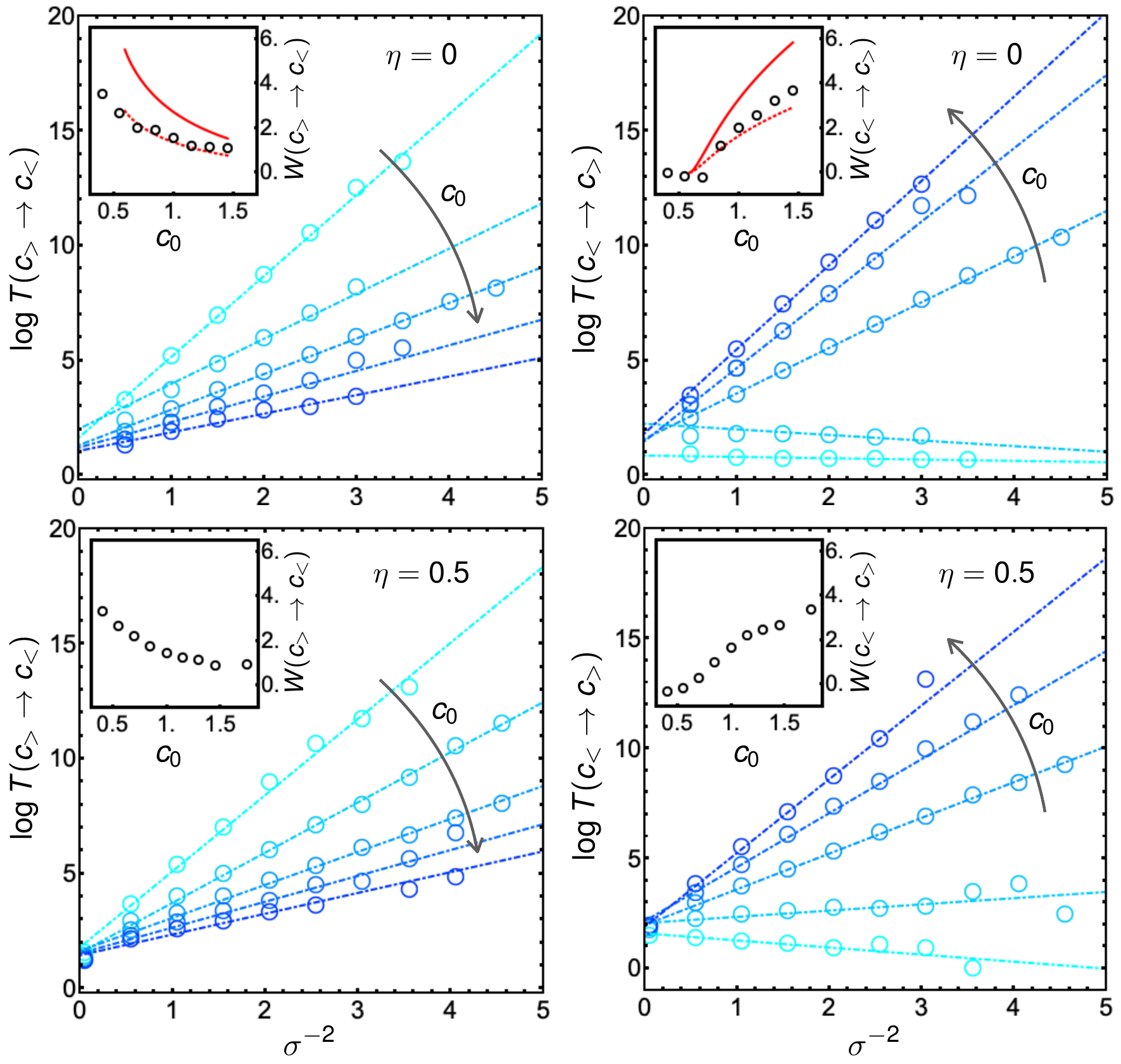}
\vspace{-0.2cm}
 \caption{Plot of  $\log T(c_> \to c_<)$ (left row) and $\log T(c_< \to c_>)$ (right row) vs $\sigma^{-2}$ for different values $c_0$, and $\eta=0$ (top graphs) and $\eta=0.5$ (bottom row). The value of $c_0$ increases with the points tonality becoming darker. The linear dependence confirms the validity of \eqref{eq:activated}. The inset shows the corresponding barriers $W$ as a function of $c_0$. For $\eta=0$, we plot the continuous time prediction \eqref{eq:kramers} with $\varepsilon=1$ (solid red), which overestimates the true barriers (dotted red) by a factor $\approx 2$.}
 \label{fig:noiseplot}
\end{figure}

It is straightforward to generalise the approach of \cite{demaeyer2013trace} and see that the jump rate has the same exponential dependence on $\sigma^2$ when the correlation time $T_\eta$ is non-zero, as confirmed by Fig \ref{fig:noiseplot}. However, finding the value of $W$ is more complicated. Approximation methods can been devised in the continuous time limit, that we describe now. 

\paragraph{Continuous Time Limit}

Let us slightly change the dynamics by assuming that $x_t$ depends not on the previous value $x_{t-1}$ but rather on an exponential moving average $\bar{x}_{t-1}$ of past values of $x$, defined recursively as
\begin{equation}\label{eq:ema}
    \bar{x}_{t-1} = (1 - \varepsilon) \bar{x}_{t-2} + \varepsilon {x_{t-1}}
 \end{equation}
 Eq. (\ref{eq:gaspard}) instead reads $x_t = H(\bar{x}_{t-1}) + \xi_t$. Eliminating $x_t$ yields
 \begin{equation}\label{eq:gaspard2}
     \bar{x}_{t} - \bar{x}_{t-1} = \varepsilon (H(\bar{x}_{t-1}) - \bar{x}_{t-1} + \xi_t),
 \end{equation}
In the limit $\varepsilon \to 0$, this equation becomes a Langevin (or SDE) equation for $\bar{x}_{t}$, for which a host of results are available. It is useful to introduce a potential function $V(x)$ such that $V'(x)=x - H(x)$.
%\textcolor{red}{I think that it might be useful to plot $-V(x)$ on the bottom panel of Fig. 1, as the equilibrium distribution for $\log p(c)$ in the continuum approximation, to make the discussion clearer to the reader. This would also to visualize in what sense the barrier is overestimated (although the simulations are not at $\eta=0$.}
% \[
% - \frac{{\rm d}V}{{\rm d}x} = H(x) - x.
% \]
%The shape of $V(x)$ for parameters corresponding to the left region of the $C$ phase (indicated by a star in Fig.~\ref{fig:parameter_space}) is shown in Fig. \red{**} \red{A faire ??}. 
The potential $V(x)$ has two minima (``valleys'') corresponding to $\log c_<$ and $\log c_>$ and a maximum (``hill'') corresponding to $\log c^*$ (see lower panels of Fig.~\ref{fig:dynamics}). With this representation, the dynamics of $\bar{x}_{t}$ under Eq. (\ref{eq:gaspard2}) becomes transparent: for  long stretches of time, $\bar{x}_{t}$ fluctuates around either $x_<=\log c_<$ or $x_>=\log c_>$, until rare fluctuations of $\xi_t$ allow the system to cross the barrier between the two valleys. Calculating the rate $\Gamma$ of these rare events is a classic problem, called the {\it Kramers problem} (for a comprehensive review, see \cite{hanggi1990reaction}). In the limit $T_\eta=0$ where the noise is white, the final exact expression is, for $\sigma \to 0$:
\begin{eqnarray}\label{eq:kramers}
     \Gamma(x_> \to x_<) &=& \frac{\sqrt{|H'(x_>)H'(x^*)|}}{2 \pi} \exp\left(-\frac{2W}{\varepsilon \sigma^2}\right),\nonumber \\ \qquad W &:=& V(x^*)-V(x_>), 
\end{eqnarray}
and {\it mutatis mutandis} for $\Gamma(x_< \to x_>)$. Such a prediction is compared with numerical simulations in Fig. \ref{fig:noiseplot}; it overestimates the real barrier by a factor $\approx 2$. The most important features is the exponential dependence of this rate on the height of the barrier $W$ and on the inverse noise variance~$\sigma^2$. The generalisation of Kramers' result for so-called coloured noise (i.e. $T_\eta > 0$) is also available, see \cite{bray1989instanton}. In this case, corrections to Eq. (\ref{eq:kramers}) can be systematically computed, but the exponential dependence of $\Gamma$ on $\sigma^{-2}$ is preserved.  

\paragraph{Exponential dependence and ``unknown knowns''}
 
It is worth emphasising the economic consequences of this exponential dependence of the probability of crises in our model. Clearly, any small uncertainty about the parameters of the model (i.e. $c_0, c_{\min}, c_{\max}, \theta$) or for that matter the precise specification of the function $G(c)$, or any other feature neglected in the model, will no doubt affect the precise value of the barrier $W$. 
%But in order to potentially describe situations like the 2008 crisis, the model must be somewhere in the $C$ phase, probably not too far from to the $C/B^+$ boundary, and with parameters such that the crisis probability is very small, i.e. in a regime where $W/\sigma^2 \gg 1$. 
%{One can imagine that the economy 2008 GFC can be pictorially described by the model for a choice of the parameter corresponding to the $C$ phase, probably not too far from to the $C/B^+$ boundary, such that the crisis probability is very small, i.e. in a regime where $W/\sigma^2 \gg 1$. Yet}
%But 
But in a regime where $W/\sigma^2 \gg 1$ any uncertainty on $W$ is exponentially amplified. Take for example $W/\sigma^2=25$; a small relative error of $10 \%$ on $W$ changes the crisis rate by one order of magnitude. Precisely as the famous butterfly effect (i.e. the exponential sensitivity on initial conditions) forbids any deterministic description of chaotic systems, the exponential dependence of the crisis rate means that this rate is, for all practical purposes, unknowable. Since the probability of rare events cannot be determined empirically, it means that no market can provide a rational valuation of the corresponding risks. This is an interesting example of ``unknown knowns'', where what may happen is known, but its probability impossible to quantify. The impossibility to price these catastrophic risks, argued to be at the heart of the excess equity risk  premium~\cite{barro2006}, would then also be responsible for the excess volatility in financial markets. 

\section{Inflation \& Narrative-Based Monetary Policy}

In the absence of frictions, the model is usually closed by assuming a Taylor rule for the interest rate, as $r_t = \Phi \pi_t - \log \beta$, with $\Phi > 1$ fixing the amplitude of the response of the Central Bank to inflation $\pi$ \cite{gali2015monetary}. Let us first assume that the crisis probability is very small, so one can linearise the Euler equation \eqref{eq:euler} with $\varsigma=1$. Solving forward in time leads to
\begin{equation}\label{eq:inflation}
    \pi_t + \frac{\kappa_>}{\Phi} (\delta_t - \delta_{t-1}) = \left(1 - \frac{\kappa_>}{\Phi}\right) \sum_{k=0}^\infty \Phi^{-k-1} \mathbb{E}_t[\delta_{t+k+1}-\delta_{t+k}],  
\end{equation}
where $\delta_t$ is the output gap defined in \eqref{eq:delta} and $\kappa_>:=3 G'(c_>) \geq 0$. In the DSGE limit $\kappa_> \to 0$ and one recovers the standard expression \cite{gali2015monetary}. The self-reflexive feedback adds a term that depends on the past output gap trend, and changes the coefficient in front of the expected future output gap variations. Interestingly, $1-\kappa_>/\Phi$ can become negative for some range of parameters. 
Accounting for crises analytically is difficult in general, since the Euler equation cannot be linearised anymore. In order to make progress, we model the dynamics as follows: with probability $p=T^{-1}(c_> \to c_<) \ll 1$ the economy crashes between $t$ and $t+1$, and with probability $1-p$ it hovers normally around $c_>$, with small fluctuations. We also assume that $\pi_t \ll 1$. Hence we approximate the right hand side of the Euler equation \eqref{eq:euler} as: 
\begin{equation}
\frac{F(c_t)}{c_>}(1 + \Phi \pi_t)\left((1-p)\mathbb{E}^>_t[(1-\pi_{t+1}-\delta_{t+1})]+p \frac{c_>}{c_<}\right),
\end{equation}
where $\mathbb{E}^>_t$ is an expectation conditional to remaining near the high output equilibrium. This eventually leads to an extra term in \eqref{eq:inflation} equal to \begin{equation}
\delta \pi_t = -\frac{p}{\Phi-1} \, \frac{c_>-c_<}{c_<}.   
\end{equation} 
As expected, anticipation of possible crises decreases inflation; provided $c_< \ll c_>$ this correction can be substantial even when $p \ll 1$. 

Our setting corresponds, up to now, to a proto-DSGE model. Including frictions (like Calvo's staggered price adjustment) would lead to a richer model, with, for example, a modified ``New Keynesian Phillips Curve''~\cite{gali2015monetary}. Of particular importance would be to include market breakdown in crises periods, i.e. allowing for $C_t \neq Y_t$: production and consumption will not match as confidence collapses. We leave these extensions for future research.

In our opinion, however, the most important aspect of our model is that it suggests alternative, behavioural tools for monetary policy, in particular in crisis time. Beyond adjusting interest rates and money supply, policy makers can use {\it narratives} to restore trust,\footnote{The importance of narratives in economics was recently stressed in \cite{shiller2017narratives}} parameterised in our model by the threshold $c_0$. If the economy lies in the neighbourhood of the $C/B^+$ phase boundary (see again Fig.~\ref{fig:parameter_space}), a mild decrease of $c_0$, engineered by the Central Bank, can help putting back the system on an even keel.

% Create the reference section using BibTeX:
%\bibliography{basename of .bib file}
%\bibliographystyle{h-physrev}

% If your first paragraph (i.e. with the \dropcap) contains a list environment (quote, quotation, theorem, definition, enumerate, itemize...), the line after the list may have some extra indentation. If this is the case, add \parshape=0 to the end of the list environment.

%
%\begin{SCfigure*}[\sidecaptionrelwidth][t]
%\centering
%\includegraphics[width=11.4cm,height=11.4cm]{frog}
%\caption{This caption would be placed at the side of the figure, rather than below it.}\label{fig:side}
%\end{SCfigure*}

\acknow{We thank A. Amstrong, R. Farmer, X. Gabaix, S. Gualdi, A. Kirman, J. Scheinkman \& F. Zamponi for many insightful discussions on these topics. This research was conducted within the \emph{Econophysics \& Complex Systems} Research Chair, under the aegis of the Fondation du Risque, the Fondation de l'Ecole polytechnique, the Ecole polytechnique and Capital Fund Management, and is part of the NIESR {\it Rebuilding Macroeconomics} initiative. MT is a member of the {\it Institut Universitaire de France}.
}

\showacknow{} % Display the acknowledgments section

\bibliography{biblio}

% Bibliography

\end{document}